\documentclass[preprint]{aastex}

\begin{document}
\title{Measurement of the Europium Isotope Ratio for the Extremely Metal-Poor, $r$-Process-Enhanced Star
CS~31082-001\footnote{Based on data collected at the Subaru
Telescope, which is operated by the National Astronomical Observatory
of Japan.}}

\author{Wako Aoki\altaffilmark{2}, Satoshi Honda\altaffilmark{2}, Timothy C. Beers\altaffilmark{3}, Christopher Sneden\altaffilmark{4}}
\altaffiltext{2}{National Astronomical Observatory, Mitaka, Tokyo, 181-8588 Japan; email: aoki.wako@nao.ac.jp, honda@optik.mtk.nao.ac.jp}
\altaffiltext{3}{Department of Physics and Astronomy, Michigan State University, East Lansing, MI 48824-1116; email: beers@pa.msu.edu}
\altaffiltext{4}{Department of Astronomy and McDonald Observatory, University of Texas, Austin, TX 78712; email: chris@verdi.as.utexas.edu}

\begin{abstract} 

We report the first measurement of the isotope fraction of europium
($^{151}$Eu and $^{153}$Eu) for the extremely metal-poor,
$r$-process-enhanced star CS~31082--001, based on high-resolution
spectra obtained with the Subaru Telescope High Dispersion
Spectrograph.  We have also obtained new measurements of this ratio
for two similar stars with previous europium isotope measurements,
CS~22892--052 and HD~115444. The measurements were made using
observations of the Eu lines in these spectra that are most
significantly affected by isotope shifts and hyperfine splitting.  The
fractions of $^{151}$Eu derived for CS~31082--001, CS~22892--052, and
HD~115444 are 0.44, 0.51, and 0.46, respectively, with uncertainties
of about $\pm 0.1$.  CS~31082--001, the first star with a meaningful
measurement of U outside of the solar system, is known to exhibit
peculiar abundance ratios between the actinide and rare-earth elements
(e.g., Th/Eu), ratios that are significantly different from those for
other stars with large excesses of $r$-process elements, such as our
two comparison objects.  Nevertheless, our analysis indicates that the
Eu isotope ratio of CS~31082--001 agrees, within the errors, with
those of other $r$-process-enhanced objects, and with that of
solar-system material. 

\end{abstract}

\keywords{Galaxy: abundances -- nuclear reactions, nucleosynthesis, abundances -- stars: abundances -- stars: Population II -- stars: individual (CS~31082--001)}

\section{Introduction}\label{sec:intro}

The chemical composition of very metal-poor ([Fe/H]$\lesssim -2.5$
\footnote{[A/B] = $\log(N_{\rm A}/N_{\rm B})- \log(N_{\rm A}/N_{\rm
B})_{\odot}$, and $\log \epsilon_{\rm A} = \log(N_{\rm A}/N_{\rm
H})+12$ for elements A and B.}) stars is expected to be determined by
a small number of nucleosynthesis events that preceded the formation
of these objects, while that of metal-rich stars like the Sun are the
result of the accumulated stellar yields throughout the long history
of the Galaxy. In recent years, abundance studies of heavy elements in
very metal-poor stars have provided quite important constraints on
models of the astrophysical neutron-capture processes, both the
$s$low-process and the $r$apid-process.  Discoveries of
$r$-process-element-enhanced, very metal-poor stars, and their
subsequent abundance analyses, have shown that the abundance pattern
of neutron-capture elements with $Z \geq 56$ agrees very well with the
$r$-process component abundances in solar-system material \citep[e.g.,
][]{sneden96, westin00}.  These results suggest that the abundance
pattern of the $r$-process component in the solar system for $Z \geq
56$ is not the result of a mixture of quite different abundance
patterns produced by individual processes, but rather, that the
patterns produced by individual processes are quite similar throughout
Galactic history.

By way of contrast, the abundances of light {\it r}-process elements
($Z<56$) of these objects are known to show a deviation from the
$r$-process component in the solar system. For instance,
\citet{sneden00} studied the abundances of neutron-capture elements,
including six elements with $40<Z<56$, of CS~22892--052 in detail, and
found a clear deviation from the scaled solar-system $r$-process
component, especially for Y, Rh, and Ag. They concluded that different
{\it r}-process sites may be responsible for the formation of the
lighter ($40<Z<56$) and heavier ($Z\geq 56$) neutron-capture elements.
A phenomenological model for these results was proposed by
\citet{wasserburg01}. 

In addition to the above studies of the {\it total} abundances of
neutron-capture elements, analyses of isotope fractions for individual
elements are also quite important for making detailed comparison with
predictions from theoretical models for the $r$-process. Europium is
one of the elements which have comparatively large isotope shifts
between the spectral lines of individual components ($^{151}$Eu and
$^{153}$Eu). \citet{sneden02} carried out an analysis of the Eu
isotope fractions for three $r$-process-element-enhanced metal-poor
stars, and showed that the fractions of $^{151}$Eu and $^{153}$Eu are
0.5, agreeing with that of solar-system material. This demonstrates
that, even at the isotopic level, there is agreement between the
abundance pattern of {\it r}-process elements in very metal-poor stars
with the $r$-process component in the solar system.

On the other hand, the extremely $r$-process-element-enhanced, very
metal-poor ([Fe/H] $= -2.9$) star CS~31082-001, first reported on by
\citet{cayrel01}, exhibits a somewhat different abundance pattern as
compared to otherwise similar stars.  The abundances of heavy
($Z\geq72$) neutron-capture elements in this star are {\it higher}
than would be predicted from the abundances of elements with $56\leq Z
\lesssim 70$, when the scaled solar-system $r$-process component
pattern is extended to the third $r$-process peak \citep{hill02}. This
suggests that the site and/or conditions under which $r$-process
nucleosynthesis of (at least) the heavy neutron-capture elements
occurs are not unique, but have some star-to-star variation.  The
theoretical implications of CS~31082-001 have been considered in
detail by \citet{qian01}, Wanajo et al. (2002), and Schatz et
al. (2002).

The agreement of the abundance pattern of CS~31082-001 with that of
the solar-system $r$-process component only appears to be valid in the
range $56 \leq Z \lesssim 70$.  One open question at this stage is
whether the isotope ratios, such as $^{151}$Eu/$^{153}$Eu, agree with
those of other $r$-process-element-enhanced stars, as well as with
that of the solar-system material. In this paper, we report the
isotope fractions of Eu derived for three stars, CS~31082-001,
CS~22892-052 and HD~115444, based on high-resolution spectra obtained
with the Subaru Telescope and the High Dispersion Spectrograph (HDS,
Noguchi et al. 2002). The latter two objects have already been studied
by \citet{sneden02}, but for comparison purposes we have carried out
an independent analysis using new spectra with higher resolution than
were previously available.

\section{Observations and Measurements}\label{sec:obs}

Observations were carried out with the HDS of the 8.2m Subaru
Telescope. The detector is a mosaic of two 4k $\times$ 2k EEV-CCD's
with 13.5~$\mu$m pixels.  Besides CS~31082-001, we observed two other
$r$-process-element-enhanced metal-poor stars, CS~22892-052
\citep{sneden96} and HD~115444 \citep{westin00}.  These two objects
have already been studied by \citet{sneden02}, and are quite useful
for comparison purposes. Details of the observations are provided in
Table \ref{tab:obs}. While the spectra of CS~22892-052 and HD~115444
were taken with a resolving power $R=90,000$, CS~31082-001 was
observed with somewhat lower resolution, $R=60,000$, due to poor
weather conditions.  Nevertheless, this resolving power is still
sufficient for the present analysis. The S/N ratios per 0.012 {\AA}
pixel at 4100 {\AA} are 110, 65, and 300 for CS~31082-001,
CS~22892-052 and HD~115444, respectively. Data reduction was performed
in the standard way within the IRAF\footnote{IRAF is distributed by
the National Optical Astronomy Observatories, which is operated by the
Association of Universities for Research in Astronomy, Inc. under
cooperative agreement with the National Science Foundation.}
environment, following procedures described by \citet{aoki02}.

For future studies on the possible binarity of these objects, which
will be investigated on the basis of observed radial-velocity
variations, Table 1 also provides (heliocentric) radial velocities
measured for our spectra. The radial velocities were measured using
clean \ion{Fe}{1} lines. The radial velocity of CS~22892-052 obtained
in our present work is +13.2 km~s$^{-1}$, and shows no significant
changes from the results of \citet{preston01}, who suggested a small
variation of the radial velocity (with a period of 120 days) for this
object. CS~31082-001 was observed in July and October 2001. The
results of the two measurements agree very well, and no variation of
the radial velocity was found with respect to the results of
\citet{hill02}. Though there is no clear evidence for binarity of
these objects from the above measurements, further long-term
monitoring of their velocities is required to obtain any definitive
conclusions.  We note that our abundance analysis for CS~31082-001 is
based only on the October 2001 spectrum, because the quality of the
data is superior to that obtained in July 2001.

\section{Analysis of Europium isotope Ratios and Results}\label{sec:ana}

The Eu isotope ratios were measured by fitting the observed spectra of
Eu lines with synthetic spectra calculated using model atmospheres
\citep{kurucz93} including the effect of the isotope shifts. We
adopted atmospheric parameters for our objects (effective temperature:
$T_{\rm eff}$, surface gravity: $\log g$, micro-turbulent velocity:
$v_{\rm micro}$, and metallicity, as represented by the iron
abundance, [Fe/H]) that are similar to those in previous work: $T_{\rm
eff}$(K) / $\log g$ /$v_{\rm micro}$(kms$^{-1}$) / [Fe/H] =
4800/1.5/1.8/$-3.0$ for CS~31082-001 \citep{cayrel01},
4750/1.3/2.3/$-3.0$ for CS~22892-052 \citep{sneden96}, and
4650/1.5/2.1/$-3.0$ for HD~115444 \citep{westin00}.  It is worth
stressing that, in contrast to studies of elemental abundances, the
analyses of isotope ratios are generally quite {\it insensitive} to
the adopted atmospheric parameters.

The effects of isotope shifts and hyperfine splitting in Eu lines have
been discussed by \citet{lawler01}. The line list produced by
\citet{sneden02}, based on \citet{lawler01}, was applied in the
present work. \citet{sneden02} analyzed the \ion{Eu}{2} lines at 3819,
3907, 4129, and 4205~{\AA}.  Unfortunately, the 3907~{\AA} line was
not covered in our CS~31082-001 spectrum due to the gap between the
two CCD's of HDS.  Instead, we analyzed the weaker \ion{Eu}{2} line at
4435~{\AA} for CS~31082-001 and CS~22892-052. As shown in the
following analysis, this line is actually quite useful for the
measurement of the isotope ratios in these extremely Eu-enhanced
stars, because the lines at 3819, 4129, and 4205~{\AA} are
sufficiently strong that they compromise the accuracy of the derived
isotope fractions, as well as the total abundance.  We note that the
4435~{\AA} line in HD~115444 is too weak, hence it was not used in the
analysis of this object. The \ion{Eu}{2} 4522~{\AA} line has an
appropriate strength for isotopic and abundance analyses, but was not
used because of severe blending with other elemental lines.

The instrumental profile of HDS, for spectra with resolving power
higher than $R \sim 90,000$, can be well-approximated by a Gaussian
shape, but this does not pertain to spectra obtained with lower
resolving powers \citep{noguchi02}.  Accordingly, we have employed
Gaussian profiles for the instrumental line broadening for calculation
of the synthetic spectra of CS~22892-052 and HD~115444, which were
observed with $R=90,000$. For the analysis of CS~31082-001, which was
observed with $R=60,000$, we measured the instrumental profile from
the Th emission lines obtained for the wavelength calibration, and
applied the resulting profile to the spectrum synthesis calculations.

The macro-turbulence for the individual stellar atmospheres was
estimated by fitting clean \ion{Fe}{1} lines with synthetic spectra
for each object, assuming Gaussian profiles.  This approximation
should be valid since the axial rotation of these (very old) red
giants is expected to be quite small.

In order to estimate the quality of the fit between observed and
synthetic spectra, we calculated the values of reduced $\chi^{2}$,
defined as:

\begin{eqnarray}
\chi_{\rm r}^{2}= \frac{1}{\nu
-1}\sum_{i}\frac{(O_{i}-C_{i})^{2}}{\sigma_{i}^{2}} ,\nonumber
\end{eqnarray}

\noindent
where $(O_{i}-C_{i})$ is the difference between the observed and
synthetic spectra at the $i$-th spectrum point \citep[e.g., ]
[]{smith01}. The quantity $\sigma_{i}$ is defined as $\sigma_{i} =
(S/N \times \sqrt{f_{i}})^{-1}$, where $S/N$ is the signal-to-noise
ratio of the continuum level, and $f_{i}$ indicates the normalized
flux at the $i$-th point ($0\leq f_{i}\leq 1$ for absorption
profiles). In the above equation, $\nu$ is the number of degrees of
freedom in the fit, and is approximately the number of data points to
which the fit is applied (20-30 pixels). Then, $\chi_{\rm r}^{2}$ is
represented as:

\begin{eqnarray}
\chi_{\rm r}^{2} \sim <\frac{(O_{i}-C_{i})^{2}}{\sigma_{i}^{2}}> . \nonumber
\end{eqnarray}

\noindent
By dividing $O-C$ by $\sigma$, the dependence of the data quality on
the depth of the absorption is accounted for. In this definition,
$\chi_{\rm r}^{2}$ is expected to be unity in the best-fit case for
data of a given $S/N$, because the $\chi_{\rm r}^{2}$ value, taking
the $S/N$ ratio at the line into consideration, cancels out the
line-to-line and object-to-object differences in data quality. In the
following presentation, however, we show the result for a fixed value
of $S/N$, in order to compare directly the goodness of the fit between
individual lines. We fix the $S/N$ to be 100, which roughly represents
the $S/N$ ratio of the CS~31082-001 spectrum.

We show examples of the observed and synthetic spectra in Figure
\ref{fig:4129A} and \ref{fig:4435A} for the \ion{Eu}{2} 4129~{\AA} and
4435~{\AA} lines, respectively. We first assumed the fraction of
$^{151}$Eu to be 0.5, and derived the total Eu abundance for each
line. Then we fitted the synthetic spectra to the observed ones for
the longer wavelength (redder) part of the spectral line, which is
insensitive to the assumed isotope ratio, by shifting the observed
spectra.  The shifts we applied are smaller than 0.01~{\AA}. These are
reasonable, as the uncertainties of the wavelength calibration of the
spectra and the absolute wavelength of the Eu lines are likely to be
in error at this level.  We then calculated the $\chi_{\rm r}^{2}$ for
the red portion of the line to determine the wavelength shifts. We
estimated the uncertainties of the wavelength determination by taking
a value where $\chi_{\rm r}^{2}$ is twice larger than the best-fit
case.

We next proceeded to the determination of the Eu isotope fractions, as
well as the total Eu abundances.  We searched for the isotope
fractions that resulted in the smallest $\chi_{\rm r}^{2}$ for a given
Eu abundance.  This analysis was carried out by changing the Eu
abundance in steps of 0.01~dex, and adopting the Eu abundances and
isotope fractions that gave the best $\chi_{\rm r}^{2}$.  We estimated
the uncertainty of the derived Eu abundances by considering the range
in abundance over which the $\chi_{\rm r}^{2}$ is twice as large as
the best-fit case.  The errors in the isotope fractions due to the
uncertainty of the total abundance of Eu were estimated from the range
of the isotope fractions that were allowed within the adopted
abundance uncertainty.  As an example of this procedure,
Figure~\ref{fig:fit} shows the values of $\chi_{\rm r}^{2}$, as a
function of $^{151}$Eu/$^{153}$Eu, for the \ion{Eu}{2} 4129~{\AA}
line.

Table~\ref{tab:results} lists the derived isotope fractions and the Eu
abundances determined by the above analysis for each line. The
$\chi_{\rm r}^{2}$ value and the fitting error ($\sigma_{\rm fit}$)
are also given. The fitting errors given here were simply estimated as
being those obtained when the $\chi_{\rm r}^{2}$ was assumed to be
twice larger than the best-fit case at the adopted Eu abundance. We
also estimated the errors due to the uncertainties of the adopted Eu
abundance, the macro-turbulent velocity, the continuum level, and the
wavelength calibration of the spectrum and the position of the Eu line
(see below). The total error ($\sigma_{\rm total}$) was estimated by
adding, in quadrature, these errors to $\sigma_{\rm fit}$, and is also
given in Table~\ref{tab:results}.

\subsection{HD~115444}

The values of $\chi_{\rm r}^{2}$ obtained from the fits of the three
Eu lines in HD~115444 are much smaller than those obtained for the
other two stars, which is expected due to the considerably higher S/N
ratio of the spectrum of this object.  The small values of $\chi_{\rm
r}^{2}$, and the agreement of the $^{151}$Eu fractions derived from
the three lines consider, indicate that the Eu isotope fraction
derived for HD~115444 is quite reliable.  The average of the
$^{151}$Eu fraction from the three lines is fr($^{151}$Eu)=0.46.

We estimated the errors in our derived isotope fractions due to
uncertainties in (1) macro-turbulent velocity ($\Delta v_{\rm
macro}$), (2) wavelength calibration and line position ($\Delta
\lambda$), (3) continuum level of the spectrum ($\Delta$(cont)), and
(4) Eu abundance adopted ($\Delta log \epsilon$ (Eu)). The uncertainty
of the Eu abundance as well as the wavelength calibration (or absolute
line position) have been estimated as noted above, and are $\Delta
log\epsilon$ (Eu)=0.02~dex and $\Delta \lambda$=0.002~{\AA},
respectively.  We assumed uncertainties for the macro-turbulent
velocity and the continuum level to be $\Delta v_{\rm
macro}=1$~km~s$^{-1}$ and $\Delta$(cont)=1\%, respectively, for this
star. These assumption are probably conservative for a spectrum with
$S/N$ = 200 $\sim 300$. The errors due to these uncertainties are
$\Delta$fr($^{151}$Eu)=$-0.07, -0.03$, and $-0.02$ for $\Delta v_{\rm
macro}$=+1.0~km~s$^{-1}$, $\Delta \lambda$=+0.002~{\AA}, and
$\Delta$(cont)=+0.01, respectively. The sense of the change in the
derived $^{151}$Eu fraction with respect to the change of the total Eu
abundance is dependent on the line under consideration.  The
uncertainties in fr($^{151}$Eu) are about 0.02 for $\Delta log
\epsilon$ (Eu)=$\pm$ 0.02~dex.

\subsection{CS~22892-052}

A similar analysis was applied to the spectrum of CS~22892-052. The
lower $S/N$ of the spectrum of this object, as compared to that of
HD~115444, results in the larger values of $\chi_{\rm r}^{2}$, and the
larger fitting errors.  In addition, there is a difficulty in the
analysis of the \ion{Eu}{2} 3819~{\AA} line. This line is quite strong
-- the central part of the absorption line is almost saturated.  As a
result, although the line {\it depth} is insensitive to the increase
of the total Eu abundance, the line {\it width} increases along with
the Eu abundance, due to the growth of weak components in the bluer
part of the line to which weak $^{151}$Eu lines primarily
contribute. This behavior is similar to that for the increase of the
fraction of $^{151}$Eu for a given Eu abundance.  As a result, the
effects of the changes of the total Eu abundance and the fraction of
$^{151}$Eu are degenerate.  We attempted to estimate the uncertainty
of the $^{151}$Eu due to that of the Eu abundance by the same manner
as other lines, and found quite large uncertainties: $\Delta$
fr($^{151}$Eu)=0.15 and $\Delta log \epsilon$(Eu)=0.10~dex. For this
reason, we decided to exclude this line for determination of the
isotope ratio, and adopted the average of the values determined from
the other three lines, fr($^{151}$Eu)=0.51, as the final result. We
note that the uncertainty of the derived $^{151}$Eu fraction due to
the uncertainty in the total Eu abundance obtained from the
\ion{Eu}{2} 4129~{\AA} line is rather large (0.09).  This is partly
due to the strength of this line, as in the case of the 3819~{\AA}
line, but the poor quality of the fit to this line also contributes.
The uncertainties arising from other factors are small ($\leq 0.04$),
even though a larger error in the continuum determination
($\Delta$(cont)=0.02) was assumed for this object, taking the low
$S/N$ ratio of the spectrum into consideration.
    
\subsection{CS~31082-001}

The \ion{Eu}{2} lines in the spectrum of CS~31082-001 are even
stronger than those in the spectrum of CS~22892-052, due to the larger
excesses of the $r$-process elements, which are boosted by about
0.4~dex relative to this star \citep{hill02}. We excluded the
3819~{\AA} line from the analysis for this star, as we did in the
analysis of CS~22892-052. The other two lines, at 4129~{\AA} and
4205~{\AA}, are also quite strong.  The uncertainties of
fr($^{151}$Eu) due to that of the total Eu abundance are 0.07; the
most important factors arise from the total errors. In this sense, the
isotope ratio derived from the \ion{Eu}{2} 4435~{\AA} line is the most
reliable, because the line strength is appropriate for the analysis of
the total abundance and isotope fractions, even though the fitting
error for this line is larger than for other lines. The uncertainties
arising from other factors are minor.

\section{Discussion and Concluding Remarks}

Figure~\ref{fig:results} shows the results for our derived $^{151}$Eu
fractions based on each line considered; the error bars indicate the
uncertainty ($\sigma_{\rm total}$). The average of the results from
individual lines are 0.46, 0.51, and 0.44 for HD~115444, CS~22892-052
and CS~31082-001, respectively.  The calculation, weighted by
$\sigma^{-1}$, alters the results by less than 0.01.  The results for
HD~115444 and CS~22892-052 show excellent agreement with those by
\citet{sneden02}, who derived fr($^{151}$Eu)=0.5$\pm$0.1 for these
objects.

The dotted line in Figure~\ref{fig:results} indicates the $^{151}$Eu
fraction in solar-system material (fr($^{151}$Eu)=0.478, Anders \&
Grevesse 1989). Since 95\% of the Eu in solar-system material is
expected to originate from the $r$-process \citep{arlandini99}, this
ratio well represents that of the $r$-process component in the solar
system\footnote{Though the $^{151}$Eu fraction of the $s$-process
component has not been constrained by observations,
\citet{arlandini99} predicted it to be 0.54, similar to that of the
$r$-process component (0.47), based on their stellar model. For this
reason, the contribution of the $s$-process nucleosynthesis is not
estimated from the Eu isotopes. However, it should be negligible in
the very metal-poor stars studied here, in particular for Eu.}. The
$^{151}$Eu fractions in our three objects, including CS~31082-001, are
{\it consistent with the solar-system value.}

One clear difference between the abundance pattern of CS~22892-052 and
that of CS~31082-001 is the abundance ratios between the actinide and
the rare-earth elements produced by $r$-process nucleosynthesis (e.g.,
the Th/Eu ratio, Hill et al. 2002).  This difference is speculated to
arise from the variety in the ratios of the neutron-to-seed-nuclei in
the $r$-process site which contributed to the abundances of heavy
elements in these objects.  Theoretical studies of $r$-process
nucleosynthesis have shown that the neutron to seed-nuclei ratio is
strongly dependent on the entropy-per-baryon ratio and the electron
fraction ($Y_{e}$) in the nucleosyhthesis site, as well as on the
dynamic time-scale of the event \citep[e.g., ]
[]{hoffman97,otsuki00,wanajo02}.  These quantities are, however, quite
difficult to estimate from theoretical studies, hence numerical
simulations of the $r$-process often treat them as free
parameters. 

In contrast to the above, the abundance ratios of the nuclei produced by the
$r$-process surrounding Eu are expected to be insensitive to these parameters,
because the nucleosynthesis paths in this mass range (i.e., $A \sim$ 150) are
quite similar in the $r$-process models which predict the production of
actinide nuclei, even though the abundances of the actinides show a
significantly large dispersion \citep[e.g., ][]{otsuki02,wanajo02}. This
prediction is supported by the similarity in the abundance patterns of the
elements with $56 \leq Z \lesssim 70$ found in extremely metal-poor stars, as
mentioned in \S 1. The result of the present study, that similar
$^{151}$Eu/$^{153}$Eu ratios are found even in the
$r$-process-element-enhanced, extremely metal-poor stars with quite different
Th/Eu ratios, is naturally explained by the above theoretical expectation. In
conclusion, our analysis of the isotope shifts appearing in a few Eu
absorption lines shows that the Eu isotope fractions in very metal-poor,
r-process-enhanced stars exhibit the value expected from standard
nucleosynthesis models, and that they are independent of the global abundance
patterns from light to heavy r-process elements.

Analyses of isotope fractions for other neutron-capture elements will
provide quite strong constraints on modeling of the {\it r}-process
nucleosynthesis. Ba and Pb are known to show rather large isotope
shifts in their absorption lines. Though the analysis for these
isotopes will prove more difficult than that of Eu, their
isotope ratios should be sensitive to related nuclear reactions.
Eu isotopes, which are (by comparison) rather easily measured from
high-resolution spectra of suitable quality, are not sensitive to these
processes. We would like to note, however, that small differences might be
expected if detailed nucleosynthesis processes are included. For instance,
processes that occur after freeze-out of the neutron-capture elements are
expected to affect the fine-structure of the abundance patterns of individual
nuclei \citep[e.g., ][]{qian97}. These effects will appear more clearly through
examination of abundances at the {\it isotopic} level, though they are likely
to be smoothed out at the {\it elemental} level.  The typical error in modern
determinations of elemental abundances is 0.1--0.2~dex ($\sim$25--50\%).  It
thus follows that, unavoidably, some dispersion ($\lesssim$ 25\%) {\it may}
exist even in the derived abundance patterns of the neutron-capture elements
with $56\leq Z
\lesssim 70$ in $r$-process-element-enhanced, metal-poor stars.  We stress that
derived isotope ratios of $r$-process elements, such as Eu, are almost
completely free from these uncertainties, even though they are more difficult
to determine and higher quality spectra are required.  The difference in the
mean $^{151}$Eu fractions between the stars CS~22892-052 and CS~31082-001 is
$\Delta$fr($^{151}$Eu)=0.07, smaller than the uncertainty of the analysis for
individual lines.  However, the fact that the derived $^{151}$Eu fractions
obtained from the three Eu lines used in the analysis of CS~31082-001 are all
lower than those obtained for CS~22892-052 may perhaps indicate a small
difference of the $^{151}$Eu fractions between the two stars.  Further
observational study of the Eu isotopes, based on higher quality spectra, for
these, and other $r$-process-enhanced, extremely metal-poor stars, is strongly
desired.

\acknowledgments

W.A. and S.H. are grateful for valuable discussions with
Drs. T. Kajino, K. Otsuki, S. Wanajo, K. Sumiyoshi, and M. Terasawa on
the modeling of {\it r}-process nucleosynthesis and interpretation of
our observational results.  T.C.B acknowledges partial support of this
work from grants AST 00-98508 and AST 00-98549, awarded by the
U.S. National Science Foundation.

\clearpage

\begin{figure}
\caption[]{Comparison of the observed spectra (dots) and synthetic
spectra (lines) for the \ion{Eu}{2} 4129~{\AA} line. The name of the
object and the adopted fr($^{151}$Eu) value are given in each
panel. The solid line shows the synthetic spectra for the adopted
fr($^{151}$Eu); the dotted and dashed lines show those for ratios
which are smaller and larger by 0.15 in fr($^{151}$Eu),
respectively. The wavelengths and relative strength of the hyperfine
components for $^{151}$Eu and $^{153}$Eu are shown in the top panel.}
\label{fig:4129A}
\end{figure} 

\begin{figure}
\caption[]{The same as Figure~\ref{fig:4129A}, but for the \ion{Eu}{2}
4435~{\AA} line.}
\label{fig:4435A}
\end{figure}

\begin{figure}
\caption[]{Reduced $\chi^{2}_{\rm r}$ as a function of the fraction of
$^{151}$Eu determined from the 4129~{\AA} line. The asterisks, open
circles, and filled circles indicate the derived results for
HD~115444, CS~22892--052, and CS~31082--001, respectively. The values
of HD~115444 and of CS~22892-052 are multiplied by 2 and 1/2,
respectively, for clarity.}
\label{fig:fit}
\end{figure}

\begin{figure}

\caption[]{Results of the $^{151}$Eu fractions and uncertainties
($\sigma_{\rm total}$). The results obtained by consideration of
\ion{Eu}{2} 3819~{\AA}, 4129~{\AA}, 4205~{\AA}, and 4435~{\AA} are
shown by squares, diamonds, open circles, and filled circles,
respectively. The dotted line is the $^{151}$Eu fraction in
solar-system material (0.478). Note that the isotopic fractions
derived for all three lines considered for CS~31082-001 are lower than
those obtained for CS~22892-052, although the error bars overlap.}

\label{fig:results}
\end{figure}

\clearpage
\begin{deluxetable}{lcccccccl}
\tablewidth{0pt}
\tablecaption{PROGRAM STARS AND OBSERVATIONS \label{tab:obs}}
\startdata
\tableline
\tableline
Star  & Wavelength range (\AA) &  Exp.$^{\rm a}$ &  Obs. Date (JD) & Radial velocity (kms$^{-1}$)  \\ 
\tableline
CS~31082-001 & 3080-3900, 3960-4780 & 200 (5) &  26 Oct., 2001 (2452208.9) & $+138.92\pm0.28$ \\
             & 3540-4350, 4440-5250 &  20 (1) &  30 July, 2001 (2452121.1) & $+138.91\pm0.30$ \\
CS~22892-052 & 3540-4350, 4440-5250 & 120 (3) &  23 July, 2001 (2452114.0) & $+13.16\pm0.36$ \\
HD115444     & 3540-4350, 4440-5250 &  30 (2) &   5 July, 2000 (2451730.8) & $-27.12\pm0.34$ \\
             & 3080-3900, 3960-4780 &  30 (2) &  16 Apr., 2001 (2452016.0) & $-27.11\pm0.28$ \\
\tableline
\enddata
~ \\

$^{\rm a}$ Total exposure time (minute) and number of exposures.

\end{deluxetable}

\begin{deluxetable}{llccccc}
\tablewidth{0pt}
\tablecaption{RESULTS \label{tab:results}}
\startdata
\tableline
\tableline
star & line  &  fr($^{151}$Eu) &  $log \epsilon$(Eu) & $\chi_{\rm r}^{2}$ & $\sigma_{\rm fit}$ & $\sigma_{\rm total}$ \\
\tableline
HD~115444    & 3819{\AA} & 0.42 & $-1.87$ & 0.59 & 0.07 & 0.11 \\
             & 4129{\AA} & 0.48 & $-1.80$ & 0.49 & 0.09 & 0.12 \\
             & 4205{\AA} & 0.49 & $-1.76$ & 0.10 & 0.04 & 0.08 \\
\tableline
CS~22892-052 & 4129{\AA} & 0.50 & $-1.01$ & 7.19 & 0.11 & 0.15 \\
             & 4205{\AA} & 0.47 & $-0.95$ & 1.50 & 0.07 & 0.09 \\
             & 4435{\AA} & 0.57 & $-0.90$ & 3.07 & 0.12 & 0.13 \\
\tableline
CS~31082-001 & 4129{\AA} & 0.43 & $-0.62$ & 2.22 & 0.06 & 0.10 \\
             & 4205{\AA} & 0.36 & $-0.62$ & 2.50 & 0.05 & 0.09 \\
             & 4435{\AA} & 0.52 & $-0.62$ & 2.51 & 0.10 & 0.11 \\
\tableline
\enddata

\end{deluxetable}

\clearpage
\plotone{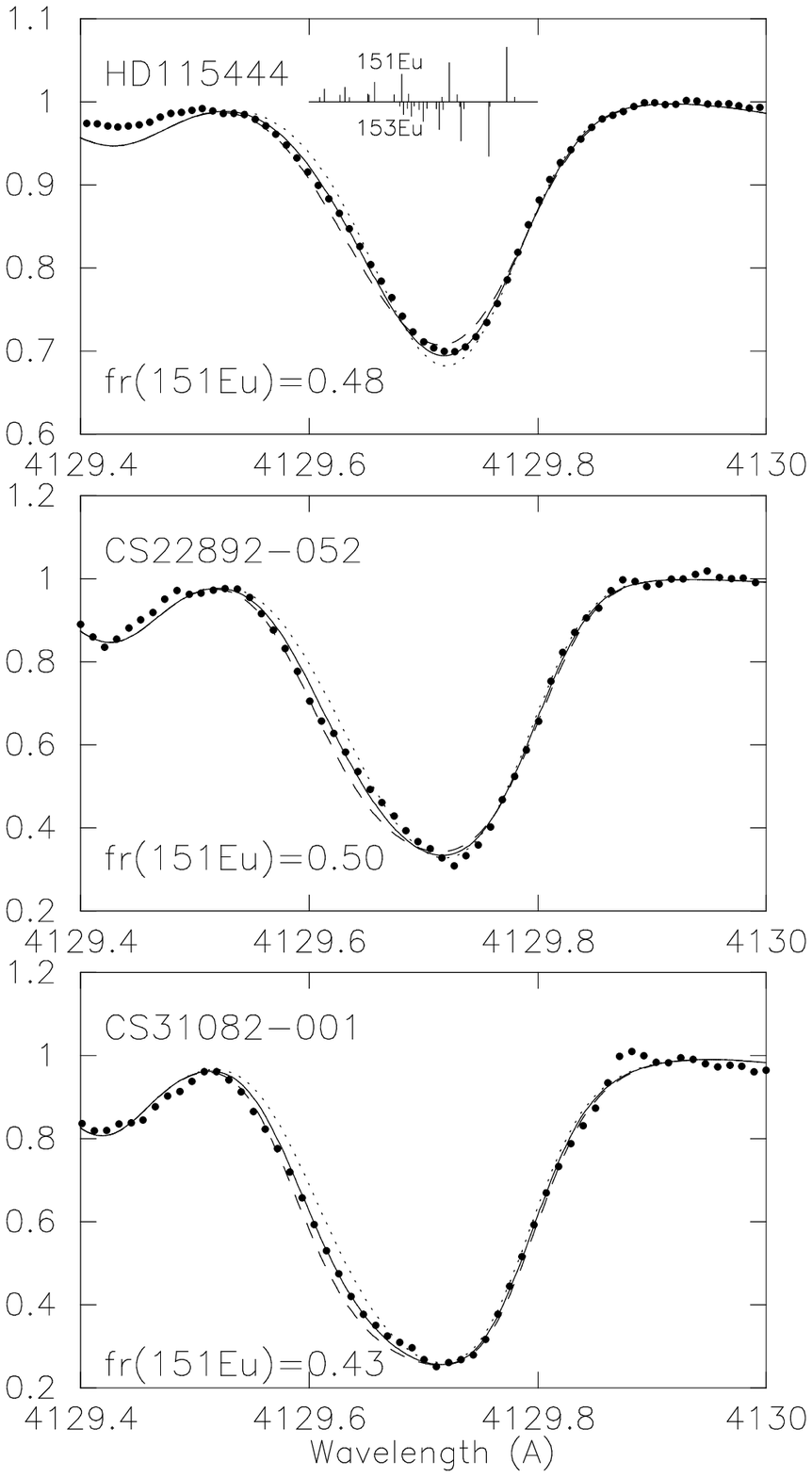}
\clearpage
\plotone{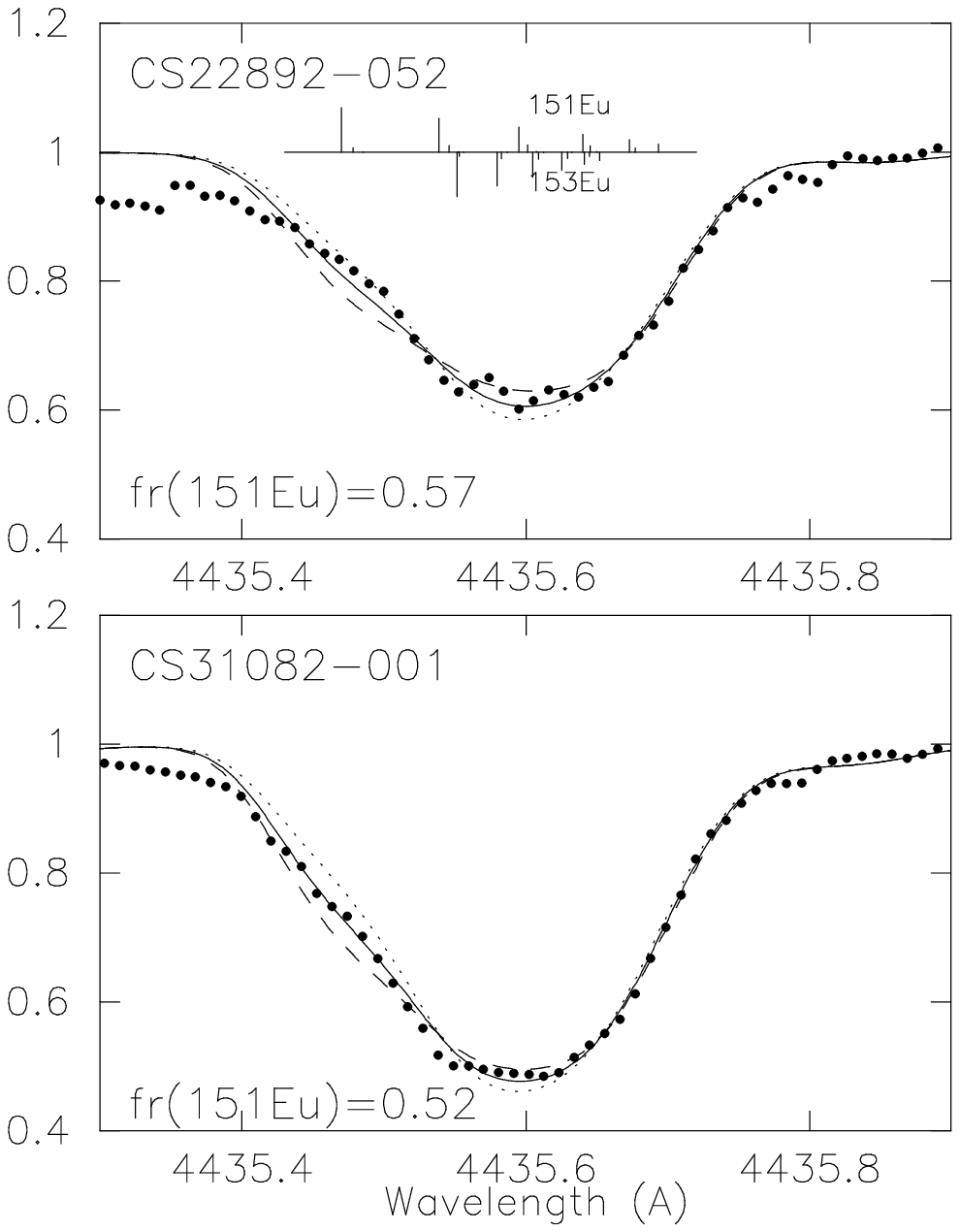}
\clearpage
\plotone{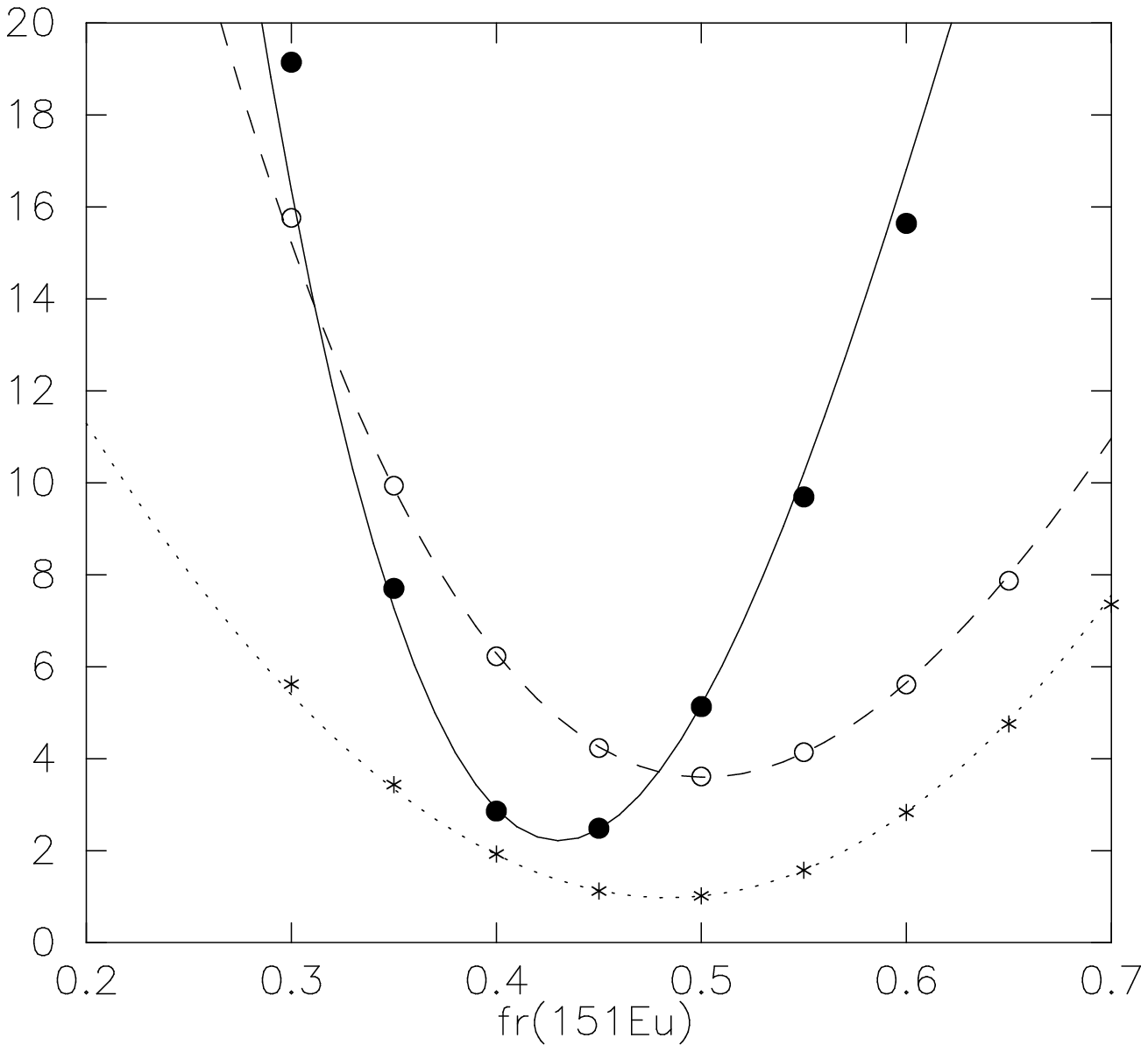}
\clearpage
\plotone{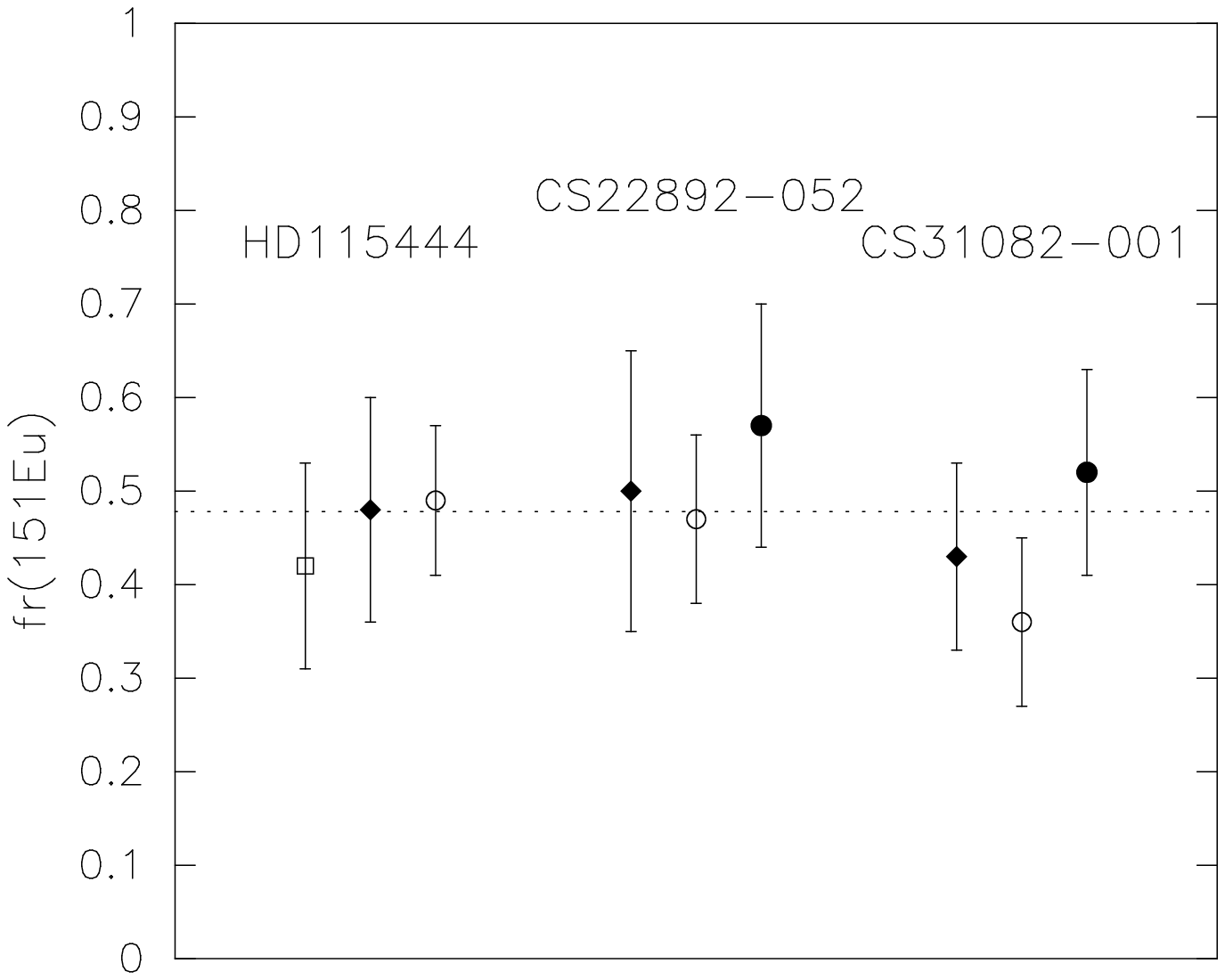}

\end{document}